\documentclass[aps,prx,superscriptaddress,reprint]{revtex4-1}
\usepackage{amsmath}
\usepackage{amssymb}
\usepackage{graphicx}
\usepackage[colorlinks,urlcolor=blue]{hyperref}
\usepackage{url}
\usepackage{hyperref}
\usepackage{color,xcolor}
\usepackage{epstopdf}
\usepackage{float}
\usepackage{ulem}
\usepackage[percent]{overpic}
\usepackage{siunitx}
\usepackage{todonotes}
\usepackage[none]{hyphenat}

\begin{document}

\title{Magnetic Correlations and Pairing Tendencies of the Hybrid Stacking Nickelate Superlattice La$_7$Ni$_5$O$_{17}$ (La$_3$Ni$_2$O$_7$/La$_4$Ni$_3$O$_{10}$) under Pressure}
\author{Yang Zhang}
\author{Ling-Fang Lin}
\email{lflin@utk.edu}
\affiliation{Department of Physics and Astronomy, University of Tennessee, Knoxville, Tennessee 37996, USA}
\author{Adriana Moreo}
\affiliation{Department of Physics and Astronomy, University of Tennessee, Knoxville, Tennessee 37996, USA}
\affiliation{Materials Science and Technology Division, Oak Ridge National Laboratory, Oak Ridge, Tennessee 37831, USA}
\author{Thomas A. Maier}
\email{maierta@ornl.gov}
\affiliation{Computational Sciences and Engineering Division, Oak Ridge National Laboratory, Oak Ridge, Tennessee 37831, USA}
\author{Elbio Dagotto}
\email{edagotto@utk.edu}
\affiliation{Department of Physics and Astronomy, University of Tennessee, Knoxville, Tennessee 37996, USA}
\affiliation{Materials Science and Technology Division, Oak Ridge National Laboratory, Oak Ridge, Tennessee 37831, USA}

\date{\today}

\begin{abstract}
Motivated by the recent rapid progress in high-$T_c$ nickelate superconductors, we comprehensively study the physical properties of the alternating bilayer trilayer stacking nickelate La$_7$Ni$_5$O$_{17}$. The high-symmetry phase of this material, without the tilting of oxygen octahedra, is not stable at ambient conditions but becomes stable under high pressure, where a small hole pocket $\gamma_0$, composed of the $d_{3z^2-r^2}$ states in the trilayer sublattice, appears. This pocket was identified in our previous work for trilayer La$_4$Ni$_3$O$_{10}$ as important to develop superconductivity. Moreover, using random-phase approximation calculations, we find a leading $s^\pm$ pairing state for the high-symmetry phase under pressure with similar pairing strength as that obtained previously for the bilayer La$_3$Ni$_2$O$_7$ compound, suggesting a similar or higher superconducting transition temperature $T_c$. In addition, we find that the dominant magnetic fluctuations in the system driving this pairing state have antiferromagnetic structure both in-plane and between the planes of the top and bottom trilayer and bilayer sublattices, while the middle trilayer is magnetically decoupled.
\end{abstract}

\maketitle

\noindent {\bf \\Introduction\\}

\begin{figure*}
\centering
\includegraphics[width=0.92\textwidth]{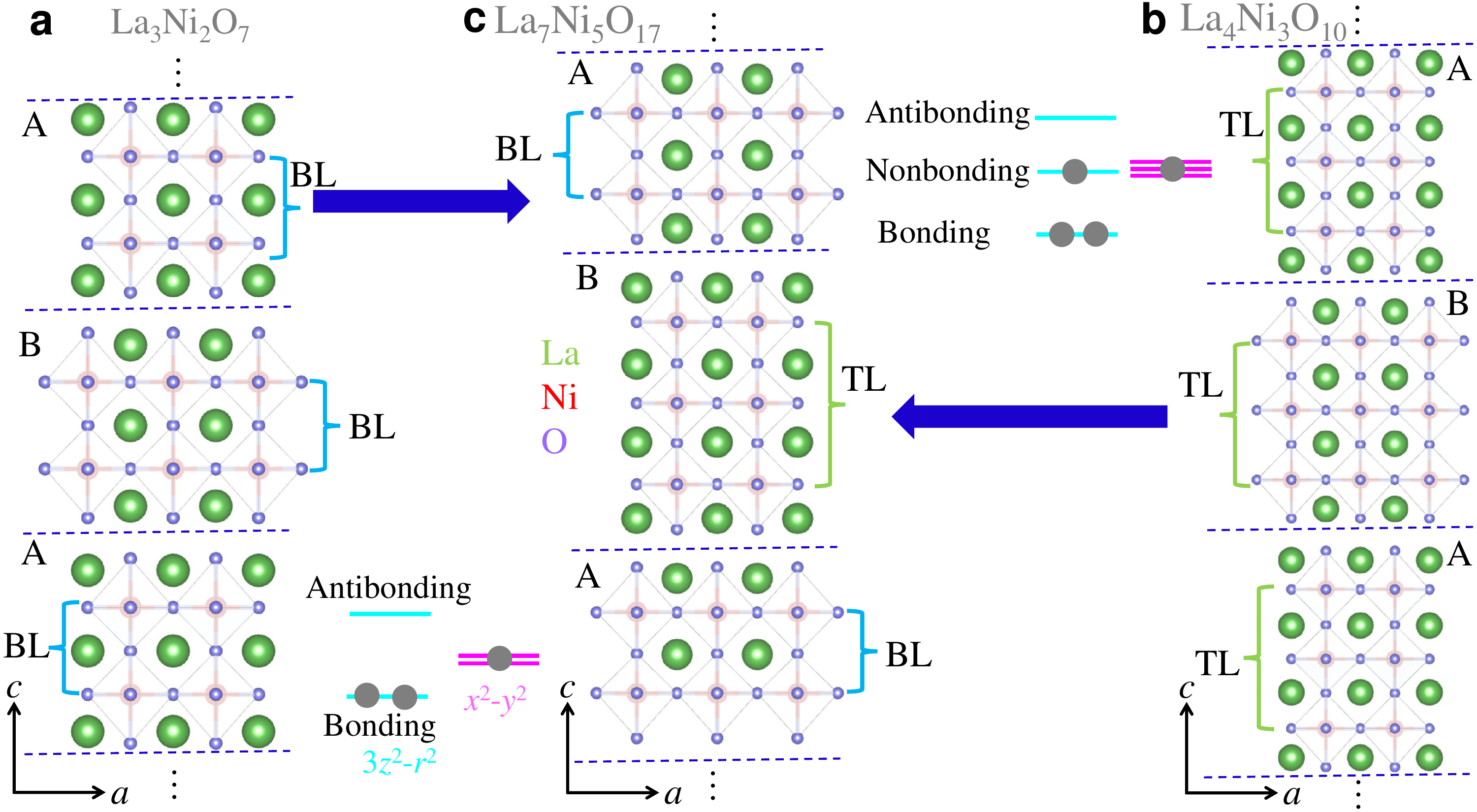}
\caption{{\bf Crystal structures and electronic configurations.} Schematic crystal structures of {\bf a} BL La$_3$Ni$_2$O$_7$, {\bf b} TL La$_4$Ni$_3$O$_{10}$, and {\bf c} the alternating BL-TL La$_7$Ni$_5$O$_{17}$ proposed here, respectively (green = La; red = Ni; purple = O). Inset: Sketches of electronic states of $e_g$ orbitals in the BL La$_3$Ni$_2$O$_7$ and TL La$_4$Ni$_3$O$_{10}$. The light blue (pink) horizontal lines represent $d_{3z^2-r^2}$ ($d_{x^2-y^2}$) states. The solid circles represent electrons. The total population $n$ of the $e_g$ electrons considered is $n = 3.0$ and $n=4.0$, for the BL with two sites and the TL with three sites, respectively. All crystal structures were visualized using the VESTA code~\cite{Momma:vesta}.}
\label{crystal}
\end{figure*}

The recent discovery of superconductivity under pressure in the Ruddlesden-Popper (RP) nickelate systems La$_3$Ni$_2$O$_7$~\cite{Sun:arxiv} and La$_4$Ni$_3$O$_{10}$~\cite{Zhu:arxiv11,Li:cpl} opened a new avenue for the study of unconventional high-{\it $T_{\rm c}$} superconductivity in the field of condensed matter physics and material sciences~\cite{LiuZhe:arxiv,Zhang:arxiv-exp,Hou:arxiv,Yang:arxiv09,Wang:arxiv9,Dong:arxiv12,Sakakibara:arxiv09,Zhang:arxiv11}. La$_3$Ni$_2$O$_7$ has the ABAB bilayer (BL) stacking structure~\cite{Zhang:jmst,Whang:jac}, where superconductivity with a high transition temperature up to 80 K~\cite{Sun:arxiv} was obtained after a first-order structural pressure-induced transition around 14 GPa~\cite{Zhang:arxiv-exp} that suppresses the tilting of the NiO$_6$ octahedra (see Fig.~\ref{crystal}{\bf a}). Very recently, J. Li et. al.~\cite{Li:arxiv24} found that superconductivity persists up to 90 GPa. Moreover, they measured the Meissner effect of the superconducting state using the $ac$ magnetic susceptibility reporting that the superconducting volume fraction is around $48 \%$~\cite{Li:arxiv24}.

La$_4$Ni$_3$O$_{10}$ also has a similar ABAB trilayer (TL) stacking structure~\cite{Zhang:nc20,Zhang:jmst,Li:arxiv11}. In this case pressure smoothly suppresses the distortion of the NiO$_6$ octahedra, yielding a high-symmetry I4/mmm phase~\cite{Zhu:arxiv11,Li:cpl}, without the tilting of oxygen octahedra, around 15 GPa (see Fig.~\ref{crystal}{\bf b}). After reaching the high symmetry phase, superconductivity was also found in a very broad pressure range although with a $T_c$ of about $20-30$ K~\cite{Sakakibara:arxiv09,Li:cpl,Zhu:arxiv11,Zhang:arxiv11}, lower than in the bilayer case. In addition, using the $ac$ magnetic susceptibility, a $\sim 80 \%$  superconducting volume fraction was reported~\cite{Li:cpl}.

Density functional theory (DFT) calculations indicate that both systems can be described using a Ni two-orbital BL or TL model~\cite{Sakakibara:arxiv09,Luo:prl23,Zhang:prb23,Zhang:arxiv24}, where the Fermi surface (FS) is composed of the Ni's $d_{x^2-y^2}$ and $d_{3z^2-r^2}$ orbitals~\cite{Luo:prl23,Zhang:prb23,Zhang:arxiv24,Li:nc,LaBollita:prb24}. In addition, the Ni $d_{3z^2-r^2}$ orbital can form a bonding-antibonding molecular-orbital state~\cite{Zhang:prb23,Zhang:arxiv24,LaBollita:prb24} in both systems, while the $d_{x^2-y^2}$ orbital is partially occupied, as shown in Fig.~\ref{crystal}. Due to the in-plane hybridization between the $e_g$ orbitals, the induced large inter-orbital hopping  leads to a noninteger population in both orbitals~\cite{Tian:prb24,Wang:prb24}. Furthermore, many interesting properties have been unveiled in both theoretical and experimental works, such as spin or charge correlations~\cite{Chen:arxiv2024,Chen:prL24,Xie:SB}, exotic superconducting pairing symmetries~\cite{Yang:prb23,Zhang:nc24,Liu:prl23,Liao:prb23,Qu:prl,Zhang:arxiv24,Yang:arxiv24}, rare-earth effects~\cite{Zhang:prb23-2,Geisler:qm}, and a prominent role of the Hund coupling~\cite{Lu:prl,Cao:prb23}.

Based on these established results, here we propose a very interesting challenge: what will happen if the BL and TL sublattices occur together in a single material? Previous studies in RP iridates reported a single-layer (SL) BL hybrid stacking superlattice structure in   Sr$_2$IrO$_4$/Sr$_3$Ir$_2$O$_7$~\cite{Kim:prr22,Gong:prl22}, as well as other RP perovskite superlattices~\cite{Zhai:nc14,Chen:nl,Yang:nl}. Furthermore, similar alternating SL-TL or SL-BL stacking structures were also discussed experimentally in nickelates, by several independent groups~\cite{Chen:jacs,Puphal:arxiv12,Wang:ic,Abadi:arxiv24,Li:prm24}. Thus, in principle, it appears possible to synthesize a BL-TL stacking structure in a real nickelate material. In this case, several interesting questions naturally arise: what are the electronic and magnetic properties of this BL-TL material? Could it become superconducting? And in this case, what is the pairing channel?

To answer those questions, we systematically investigate the alternating BL and TL hybrid stacking nickelate superlattice La$_7$Ni$_5$O$_{17}$ (La$_3$Ni$_2$O$_{7}$/La$_4$Ni$_3$O$_{10}$) by using DFT and random phase approximation (RPA) calculations. At ambient conditions,
the high-symmetry phase of La$_7$Ni$_5$O$_{17}$ without the tilting of oxygen octahedra is not stable it becomes stable under pressure. Similarly to other RP nickelates, La$_7$Ni$_5$O$_{17}$ also has a large transfer gap and bond-antibonding splittings of $d_{3z^2-r^2}$ orbitals, where
its electronic structureresemble the combinations of BL and TL nickelates.  Furthermore, iyr RPA calculations suggest a leading $s^\pm$ pairing state with sizable pairing strength $\lambda$ similar to that found for the pure BL nickelates, suggesting a similar or higher superconducting transition
temperature. Considering previous theoretical and experimental studies on RP nickelate superconductors, we, therefore, believe that superconductivity could also be expected to exist in La$_7$Ni$_5$O$_{17}$ for a broad pressure region as well. Moreover, we also find that the dominant magnetic fluctuations driving this pairing state have antiferromagnetic structure both in-plane and between the planes of the top and bottom TL and BL sublattices, while the middle TL layer is magnetically decoupled. However, the antiferromagnetic correlations are much weaker in the BL sublattice, compared to that in the TL sublattice.

\noindent {\bf \\Results\\}
\noindent {\small \bf \\Structural instability\\}
Considering previous work on nickelate superconductors, it has been established that superconductivity exists in the high symmetry phase without the tilting of oxygen octahedra. For this reason, we used the P4/mmm symmetry (No.123) of the high-pressure phase of the SL-TL stacking nickelate~\cite{Puphal:arxiv12}, without any tilting, to construct the BL-TL stacking structure, as shown in Fig.~\ref{crystal}{\bf c}. This corresponds to the chemical formula La$_7$Ni$_5$O$_{17}$. To better understand the physical properties of this BL-TL La$_7$Ni$_5$O$_{17}$ compound, we optimized the crystal structure for different pressures by using first-principles DFT calculations~\cite{Kresse:Prb,Kresse:Prb96,Blochl:Prb,Perdew:Prl}. Using the fully optimized crystal lattice, and a high accuracy tolerance, the space group becomes P4mm (No. 99). Because the distortion from P4/mmm (No. 123) to P4mm (No. 99) is very small, leading to a tiny difference in enthalpy ($\sim 0.05$ meV/Ni), the electronic structures of both symmetries are essentially identical (see Supplementary Note I.). This is quite similar to a previous study in the BL La$_3$Ni$_2$O$_7$ context, where both lower Fmmm and higher I4/mmm symmetries give the same results~\cite{Sakakibara:prl24} for the same reason: only a tiny distortion exists between the I4/mmm and Fmmm phases of La$_3$Ni$_2$O$_7$~\cite{Zhang:1313}.

\begin{figure*}
\centering
\includegraphics[width=0.92\textwidth]{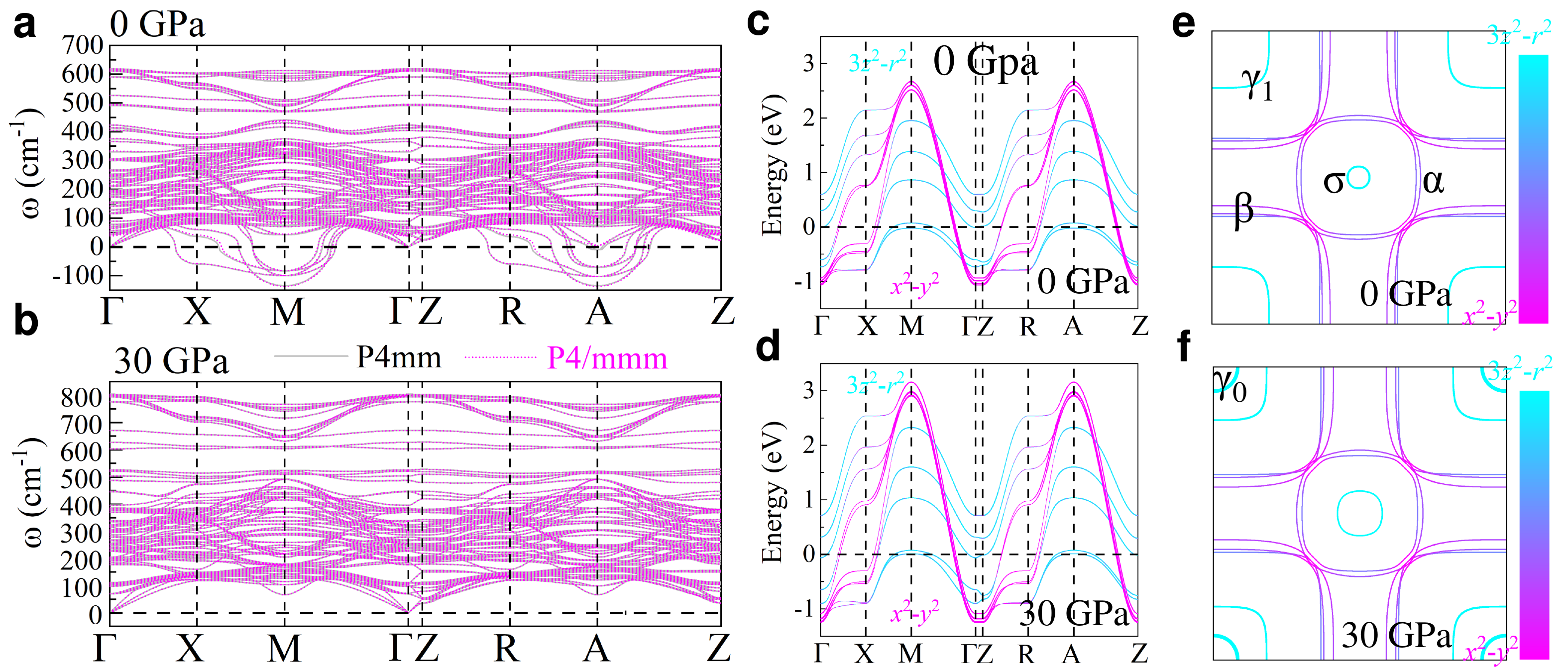}
\caption{{\bf Phonon spectrum and electronic structures.} {\bf a-b} Phonon spectrum of the BL-TL stacking La$_7$Ni$_5$O$_{17}$ compound in the P4mm (No. 99) and P4/mmm (No. 123) phases at 0 GPa, and at 30 GPa, respectively. The results of other pressures can be found in Supplementary Note II. The coordinates of the high-symmetry points in the Brillouin zone are $\Gamma$ = (0, 0, 0), X = (0, 0.5, 0), M = (0.5, 0.5, 0), Z = (0, 0, 0.5), R = (0, 0.5, 0.5) and A = (0.5, 0.5, 0.5). Band structures and FSs for {\bf c-d} 0 and {\bf e-f} 30 GPa, respectively. Here, a ten-band $e_g$ orbital tight-binding model was considered including long-range hoppings linking the BL and TL substructures, with an overall filling of $n = 7$ (i.e. 1.4 electrons per site).}
\label{Fig2}
\end{figure*}

Next, we study the structural stability of this BL-TL stacking structure. For this purpose, we calculated the phonon spectrum of the P4mm and P4/mmm phases of La$_7$Ni$_5$O$_{17}$ both without and with pressure, by using the density functional perturbation theory approach~\cite{Baroni:Prl,Gonze:Pra1,Gonze:Pra2} analyzed by the PHONONPY software~\cite{Chaput:prb,Togo:sm}. The phonon dispersion spectrum displays imaginary frequencies appearing at the high symmetry points of La$_7$Ni$_5$O$_{17}$, as displayed in Fig.~\ref{Fig2}{\bf a}, suggesting that the P4mm or P4/mmm phase is not stable at 0 GPa. However, both of them become stable under high pressure, namely without showing any imaginary frequency (see the results at 30 GPa in Fig.~\ref{Fig2}{\bf b}. This is similar to previous well-discussed nickelate superconductors~\cite{Zhang:nc24,Zhang:arxiv24}, where the untilted structure is not stable at ambient pressure but becomes stable at high pressure.

\noindent {\small \bf \\Eletronic structures\\}
Let us now discuss the electronic properties of the BL-TL stacking La$_7$Ni$_5$O$_{17}$. Similar to other nickelates~\cite{Nomura:rpp,Zhang:prb20}, the states of La$_7$Ni$_5$O$_{17}$ near the Fermi level mainly arise from the Ni $e_g$ orbitals while the O $2p$ orbitals are located deeper in energy, leading to a large charge-transfer gap between $3d$ and $2p$ states. We then constructed a ten-band ${e_g}$-orbital tight-binding model for the BL plus TL stacking structure of La$_7$Ni$_5$O$_{17}$ at both 0 GPa and 30 GPa. Here, the overall electron filling is $n = 7$, the ten bands arise from five Ni atoms in a unit cell, each contributing two orbitals, and the model includes the longer-range hoppings linking the BL and TL sublattices obtained from the maximally localized Wannier functions (MLWFs) method based on the WANNIER90 package~\cite{Mostofi:cpc}. Formally, the model's Hamiltonian is written as

\begin{eqnarray}\label{eq:H}
H_k = \sum_{\substack{i\sigma\\\vec{\alpha}\gamma\gamma'}}t_{\gamma\gamma'}^{\vec{\alpha}}
(c^{\dagger}_{i\sigma\gamma}c^{\phantom\dagger}_{i+\vec{\alpha}\sigma\gamma'}+H.c.)+ \sum_{i\gamma\sigma} \Delta_{\gamma} n_{i\gamma\sigma}\,.
\end{eqnarray}
Here, the first term represents the hopping of an electron from orbital $\gamma'$ at site $i+\vec{\alpha}$ to orbital $\gamma$ at site $i$. $c^{\dagger}_{i\sigma\gamma}$($c^{\phantom\dagger}_{i\sigma\gamma}$) is the standard creation (annihilation) operator, $\gamma$ and $\gamma'$ represent different orbitals, and $\sigma$ is the $z$-axis spin projection. $\Delta_{\gamma}$ represents the crystal-field splitting of each orbital $\gamma$. The vectors $\vec{\alpha}$ are along the three directions, defining different hopping neighbors (the entire hopping files are in the Supplementary Material).

As shown in Figs.~\ref{Fig2}{\bf c-d}, the electronic structure of La$_7$Ni$_5$O$_{17}$ appears to arise from a combined behavior between La$_3$Ni$_2$O$_7$ (BL) and La$_4$Ni$_3$O$_{10}$ (TL): the $d_{3z^2-r^2}$ orbitals display the bonding-antibonding, or the bonding-antibonding-nonbonding splitting behavior characteristic of the BL and TL sublattices, respectively, while the $d_{x^2-y^2}$ orbital remains decoupled among the planes. The bandwidth of the $e_g$ orbitals increases by $\sim  20\%$ from 0 GPa to 30 Gpa, suggesting an enhancement of itinerant properties under pressure. However, the TL bonding state involving the $d_{3z^2-r^2}$ orbital does not touch the Fermi level at 0 GPa, but crosses the Fermi level at 30 GPa. This results in a small hole pocket $\gamma_0$ in La$_7$Ni$_5$O$_{17}$ at 30 GPa, which is {\it absent} at 0 GPa, as displayed in Figs.~\ref{Fig2}{\bf e-f}. This small hole pocket from the TL sublattice is important for the superconductivity as argued in the following discussion, and as found in our previous literature on the TL La$_4$Ni$_3$O$_{10}$~\cite{Zhang:arxiv24}. There is a larger hole $\gamma_1$ pocket composed by the $d_{3z^2-r^2}$ orbitals from the BL sublattice as well, but we find that it does not play a role in pairing. Finally, the $\alpha$ and $\beta$ sheets of the Fermi surface have a multilayer splitting behavior, as displayed in Figs.~\ref{Fig2}{\bf e-f}.

\begin{figure*}
\centering
\includegraphics[width=0.92\textwidth]{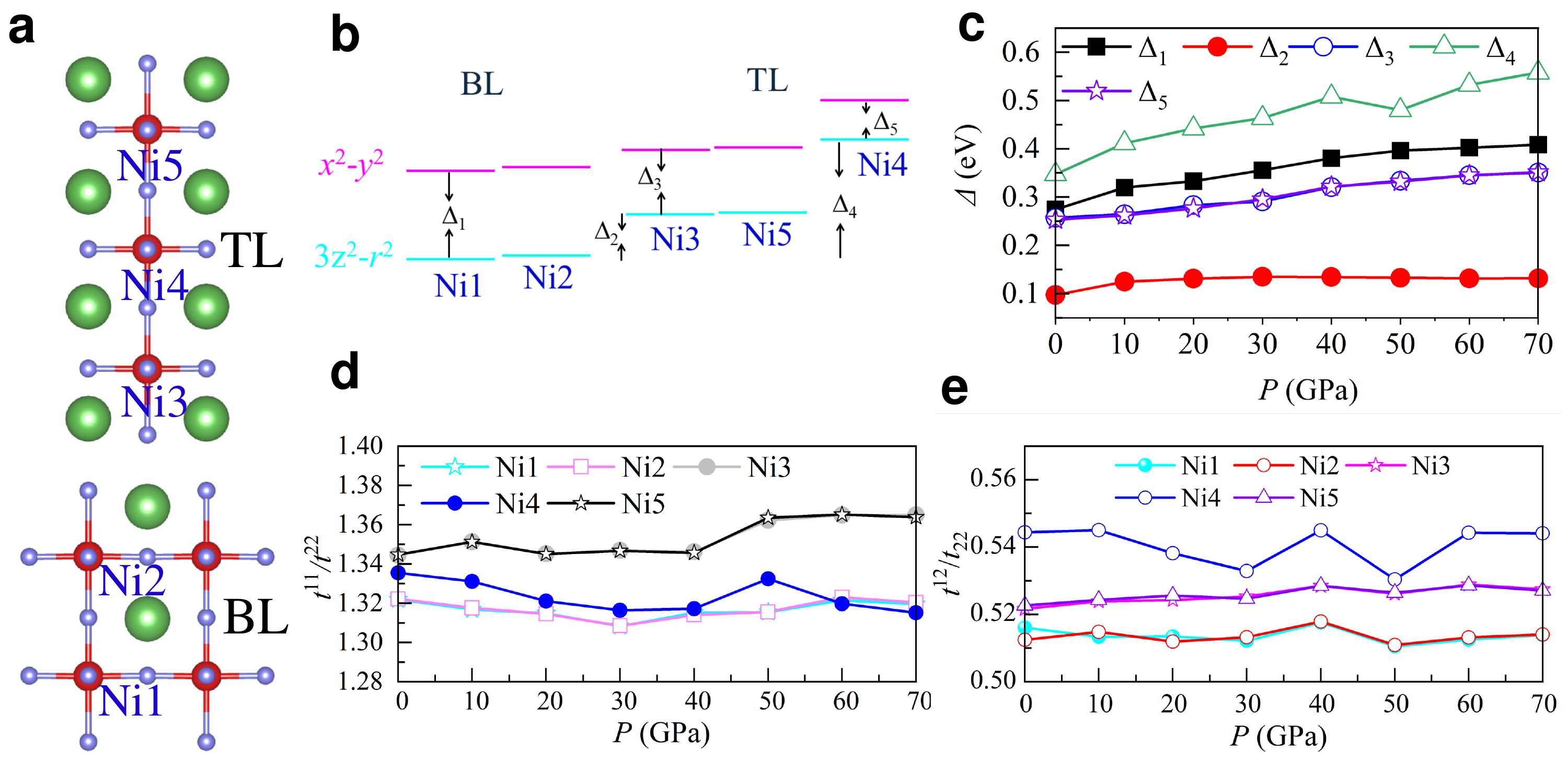}
\caption{{\bf Pressure effect.} {\bf a} Schematic crystal structure of the BL-TL stacking of La$_7$Ni$_5$O$_{17}$, where the different Ni sites are marked. {\bf b} Crude sketches of the crystal-field splitting of the $e_g$ orbitals for different Ni sites. {\bf c} Crystal-field splitting energies as a function of pressure. {\bf d} Ratio of hoppings $t_{z}^{11}$/$t_{x/y}^{22}$, and {\bf e} ratio of hoppings $t_{x/y}^{12}$/$t_{x/y}^{22}$, vs. pressure. The $\gamma = 1$ and $\gamma = 2$ orbitals correspond to the $d_{3z^2-r^2}$ and $d_{x^2-y^2}$ orbitals, respectively.}
\label{hopping}
\end{figure*}

Different from the SL-TL stacking nickelate (SL: $n = 1.77$ and TL: $n = 4.23$ )~\cite{Zhang:1313}, we do not observe a robust ``charge transfer'' effect between the BL and TL layers at both 0 GPa (BL: $n = 2.96$ and TL: $n = 4.04$) and at 30 Gpa (BL: $n = 3.00$ and TL: $n = 4.00$), respectively. In addition, both Ni sites of the top and bottom layers of the BL and TL sublattices (see Fig.~\ref{hopping}{\bf a}) have a very small on-site energy difference ($< 1$ meV), as well as a very small hopping difference between Ni3-Ni4 and Ni4-Ni5 ($< 1$ meV). Furthermore, the in-plane hybridization ($t_{x/y}^{12}$/$t_{x/y}^{22}$) between the $d_{3z^2-r^2}$ and $d_{x^2-y^2}$ orbitals is slightly enhanced in the BL sublattice ($\sim 0.512$) in La$_7$Ni$_5$O$_{17}$, compared with that in the BL La$_3$Ni$_2$O$_7$ ($\sim 0.475$).

\noindent {\small \bf \\Pressure effect\\}
As displayed in Fig.~\ref{hopping}{\bf b}, the $d_{3z^2-r^2}$ orbitals from the BL sublattice have slightly lower on-site energy than that of the bottom and top layers from the TL sublattice, while the  $d_{3z^2-r^2}$ orbitals of the Ni4 site from the middle layer of the sublattice has much higher on-site energy. In addition, we also show several key crystal-field splitting energies for different pressures in Fig.~\ref{hopping}{\bf c}. As pressure increases, the values of the crystal splittings between the $e_g$ orbitals ($\Delta_1$, $\Delta_3$ and $\Delta_5$) also increase, as do the values of both the interorbital and intraorbital hopping matrix elements. However, the hopping ratios $t_z^{11}/t_{x/y}^{22}$ for the nearest-neighbor Ni-Ni along the interlayer direction do not change much for different Ni sites under pressure, as shown in Fig.~\ref{hopping}{\bf d}. Moreover, the different ratios of in-plane hybridizations ($t_{x/y}^{12}$/$t_{x/y}^{22}$) of $e_g$ orbitals do not change much either, but this hybridization is stronger in the TL sublattice than that in the BL sublattice, at least in the pressure region we studied (see Fig.~\ref{hopping}{\bf e}).

\noindent {\small \bf \\Pairing symmetry under pressure\\}
Next, we discuss the possibility of superconductivity under pressure in the La$_7$Ni$_5$O$_{17}$ system of our focus. To study the superconducting pairing tendencies in this BL-TL stacking La$_7$Ni$_5$O$_{17}$, we have used multi-orbital RPA calculations for the tight-binding model in Eq.~(\ref{eq:H}) including the SL and TL sublattices by using hopping and crystal-field splittings obtained for 30 GPa. The RPA calculations of the pairing vertex are based on a perturbative weak-coupling expansion in the local Coulomb interaction matrix, which contains intra-orbital ($U$), inter-orbital ($U'$), Hund's rule coupling ($J$), and pair-hopping ($J'$) terms~\cite{Kubo2007,Graser2009,Altmeyer2016,Romer2020}. The pairing strength $\lambda_\alpha$ for the pairing channel $\alpha$, and its corresponding pairing structure $g_\alpha({\bf k})$, are obtained by solving an eigenvalue problem of the form
\begin{eqnarray}
\int_{FS} d{\bf k'} \, \Gamma({\bf k -  k'}) g_\alpha({\bf k'}) = \lambda_\alpha g_\alpha({\bf k}),\,
\label{eq:pp}
\end{eqnarray}
where the momenta ${\bf k}$ and ${\bf k'}$ are restricted to the FS, and $\Gamma({\bf k - k'})$ is the irreducible particle-particle vertex. In the RPA approximation, the dominant term entering $\Gamma({\bf k-k'})$ is the RPA spin susceptibility tensor $\chi^s_{\ell_1\ell_2\ell_3\ell_4}({\bf k-k'})$, where $\{\ell_i\}$ denote the ten Ni-$d$ orbitals of the BL-TL structure shown in Fig.~3{\bf a}. Here we have used $U=0.8$ eV, $U'=U/2$, $J=J'=U/4$, as in previous literature to facilitate comparison and study trends.

\begin{figure}
\centering
\includegraphics[width=0.48\textwidth]{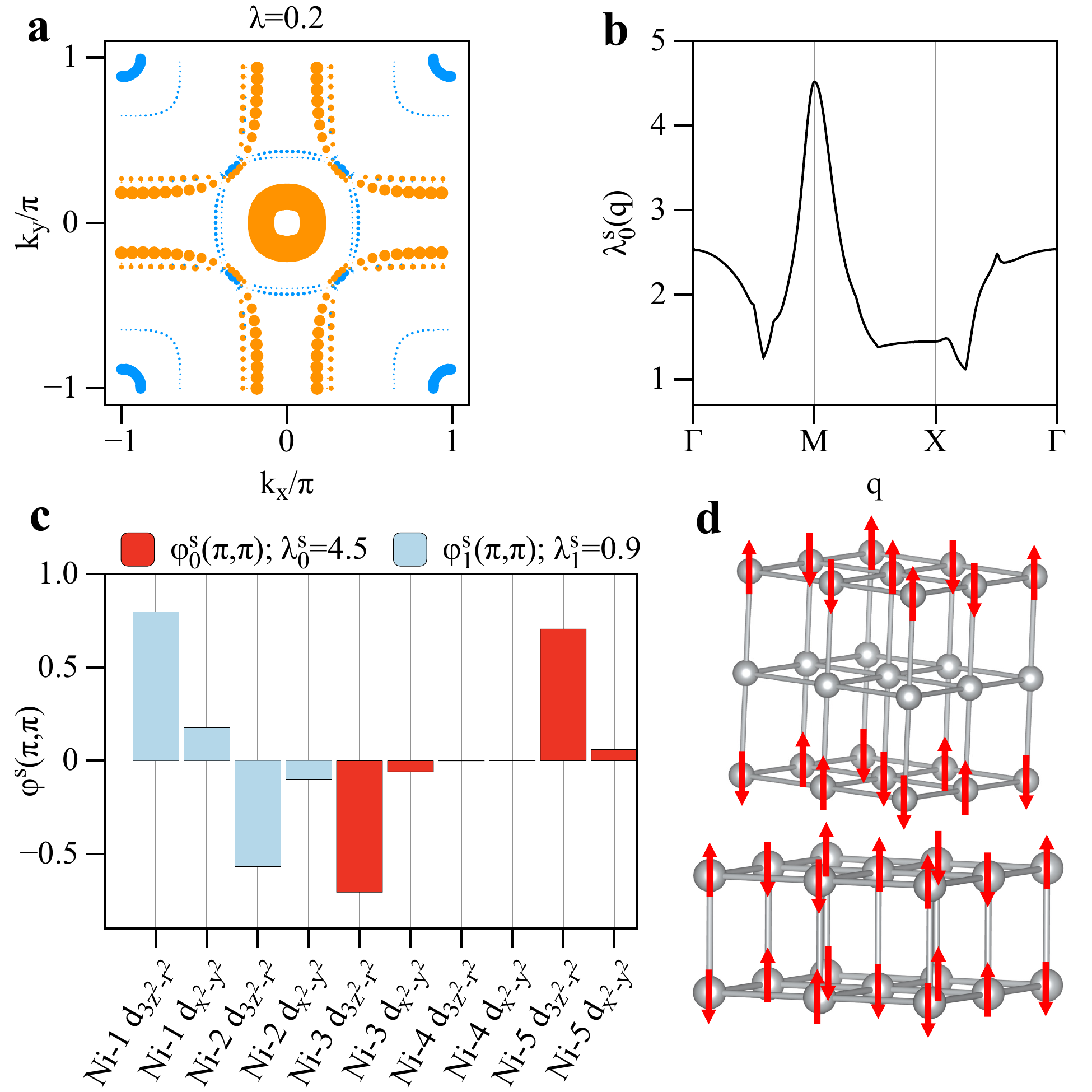}
\caption{{\bf Superconducting pairing symmetry and magnetic correlations.} {\bf a} The RPA calculated leading superconducting singlet gap structure $g({\bf k})$ for momenta ${\bf k}$ on the FS of La$_7$Ni$_5$O$_{17}$ with pairing strength $\lambda=0.2$ at 30 GPa for $n = 7.0$. The sign of $g_\alpha({\bf k})$ is indicated by the color (orange = positive, blue = negative), and its amplitude by the point sizes. Here we used Coulomb parameters $U=0.8$ eV, $U'=U/2$, and $J=J'=U/4$, and the calculation was performed for a temperature of $T=0.01$ eV.
{\bf b} Leading eigenvalue $\lambda^s_0({\bf q})$ of the spin susceptibility tensor $\chi^s_{\ell_1\ell_1\ell_2\ell_2}({\bf q})$, where $\ell_i$ are the ten Ni-$d$ orbitals included in the model, along a high-symmetry path in the Brillouin zone. {\bf c} Leading and sub-leading eigenvectors $\varphi^s_\alpha(\pi,\pi)$ of $\chi^s_{\ell_1\ell_1\ell_2\ell_2}({\bf q})$ for its maximum at ${\bf q}=(\pi,\pi)$ with corresponding eigenvalues $\lambda^s_0=4.5$ and $\lambda^s_1=0.9$, respectively. {\bf d} Schematic of the magnetic structure of the BL-TL La$_7$Ni$_5$O$_{17}$, where spin-up and spin-down are marked by red arrows. In the middle layer of the TL sublattice, the spin is zero.}
\label{Pairing}
\end{figure}

As shown in Fig.~\ref{Pairing}{\bf a}, we find that the leading eigenvector $g_0({\bf k})$ of the gap equation (\ref{eq:pp}) has a sign-switched $s^\pm$ structure and eigenvalue $\lambda_s=0.2$. This calculation was performed for the model in Eq.~(\ref{eq:H}) for the BL-TL stacking La$_7$Ni$_5$O$_{17}$ with overall filling $n = 7.0$ at 30 GPa and a temperature $T=0.01$ eV. The gap is largest on the inner $\Gamma$ pocket $\sigma$ and the order parameter has opposite sign on the small M-centered pocket $\gamma_0$. We find that this leading instability is well separated from the next leading gap structures that have an order of magnitude smaller eigenvalues. For $T=0.01$ eV, the pairing strength $\lambda_s=0.20$ is already larger than the pairing strentgh $\lambda_s=0.14$ we obtained for the bilayer 327-LNO at 25 GPa and the same interaction strength, but at lower $T=0.002$ eV \cite{Zhang:nc24}, suggesting that the pairing correlations in the BL-TL stacking system are stronger and could potentially lead to higher $T_c$ than in the bilayer.

For comparison, we do not find any singlet pairing solution with sizable $\lambda$ for the system at 0 GPa, suggesting the importance of pressure effects for superconductivity. Moreover, the FSs of other pressures look quite similar to the 30 GPa in the whole pressure range (see Supplementary Note III.) for which the high-symmetry phase is stable (up to 70 GPa). Thus, it is reasonable to expect that superconductivity can exist over a broad pressure range, considering previous studies of nickelate superconductors~\cite{Sun:arxiv}.

\noindent {\small \bf \\Magnetic correlations under pressure\\}
Finally, to study the spin fluctuations driving the leading pairing instability in the BL-TL stacking La$_7$Ni$_5$O$_{17}$ system, we analyze the RPA enhanced spin susceptibility tensor $\chi({\bf q}, \omega=0)$ that is obtained from the Lindhart function tensor $\chi_0({\bf q})$ as
\begin{eqnarray}
\chi({\bf q}) = \chi_0({\bf q})[1-{\cal U}\chi_0({\bf q})]^{-1}.
\end{eqnarray}
Here, all the quantities are rank-four tensors in the orbital indices $\ell_1, \ell_2, \ell_3, \ell_4$ and ${\cal U}$ is a tensor involving the interaction parameters~\cite{Graser2009}. The physical spin susceptibility is otained by summing the pairwise diagonal $\chi_{\ell_1\ell_1\ell_2\ell_2}({\bf q})$ over $\ell_1$, $\ell_2$.

Fig.~\ref{Pairing}{\bf b} plots the leading eigenvalue $\lambda^s_0({\bf q})$ of $\chi_{\ell_1\ell_1\ell_2\ell_2}({\bf q})$ for momenta ${\bf q}$ along high-symmetry directions in the Brillouin zone. The strong peak at ${\bf q}=(\pi,\pi)$ (M-point) indicates strong in-plane antiferromagnetic correlations. Focusing on the corresponding eigenvector $\varphi^s_0({\bf q})$ (red bars in panel {\bf c}) for the maximum at ${\bf q}=(\pi,\pi)$, one sees the the antiferromagnetic $(\pi,\pi)$ peak mainly arises from scattering between the Ni-3 and Ni-5 $d_{3z^2-r^2}$ states in the TL subsystem. The magnetic correlations are antiferromagnetic between the top and bottom layer, due to the opposite sign between the Ni-3 and Ni-5 sites, while the middle layer does not contribute to the $(\pi,\pi)$ correlations. To assess the correlations in the bilayer subsystem, we also plot the sub-leading eigenvector $\varphi^s_1(\pi,\pi)$ (blue bars in panel {\bf c} of Fig.~\ref{Pairing}). This eigenvector has opposite contributions from the bottom (Ni-1) and top (Ni-2) $d_{3z^2-r^2}$ states and a much smaller eigenvalue $\lambda^s_1=0.9$ than the leading eigenvector ($\lambda^s_0=4.5$), indicating much weaker antiferromagnetic correlations in the BL subsystem. The resulting magnetic correlations are illustrated in panel {\bf d} of Fig.~\ref{Pairing}, showing that the top and bottom layers are antiferromagnetically correlated both in-plane and between the
planes in both the BL and TL sublattices, while the middle layer of the TL sublattice has zero spin density.

Previously, for the pure BL La$_3$Ni$_2$O$_7$ system, we found that the peak in the magnetic susceptibility was near $q = (\pi, 0)$ or $(0, \pi)$ rather than at $(\pi,\pi)$. This difference can be intuitively understood. Compared to the pure BL stacking case with orbital site occupations of 0.9 for $d_{3z^2-r^2}$ and 0.6 for $d_{x^2-y^2}$, the electronic densities are 0.85 and 0.65 per site for the $d_{3z^2-r^2}$ and $d_{x^2-y^2}$ orbitals in the BL sublattice of the BL-TL stacking La$_7$Ni$_5$O$_{17}$, respectively. Thus, it is reasonable to expect that the ferromagnetic tendency along one in-plane direction due to the recently proposed half-empty mechanism~\cite{Lin:prl21,Lin:cp} should be reduced, leading to the G-AFM state despite the filling of 1.5 electrons per site in the BL sublattice.

\noindent {\bf \\Conclusion\\}

In summary, we have systematically studied the alternating BL-TL stacking nickelate superlattice La$_7$Ni$_5$O$_{17}$ (La$_3$Ni$_2$O$_{7}$/La$_4$Ni$_3$O$_{10}$) by using DFT and RPA calculations. Similarly to other RP nickelates,
the high-symmetry phase of La$_7$Ni$_5$O$_{17}$ without the tilting of oxygen octahedra is not stable at ambient conditions but becomes stable under pressure. The electronic structure of this BL-TL stacking La$_7$Ni$_5$O$_{17}$ resemble the combinations of BL and TL nickelates.  Based on RPA calculations, we obtain a leading $s^\pm$ pairing state  with sizable pairing strength $\lambda$ similar to that found for the pure BL nickelates. Considering previous theoretical work on nickelate superconductors, we therefore predict that superconductivity should also be expected to exist in La$_7$Ni$_5$O$_{17}$ for a broad pressure region. Moreover, we find that the dominant magnetic fluctuations driving this pairing state have antiferromagnetic structure both in-plane and between the planes of the top and bottom TL and BL sublattices, while the middle TL layer is magnetically decoupled. However, the antiferromagnetic correlations are much weaker in the BL sublattice, compared to that in the TL sublattice.  To obtain this alternating BL-TL stacking nickelate superlattice La$_7$Ni$_5$O$_{17}$, a possible way is to grow a sample layer-by-layer by using molecular beam epitaxy growth, as obtained in SL-BL stacking iridate Sr$_2$IrO$_4$/Sr$_3$Ir$_2$O$_7$~\cite{Kim:prr22,Gong:prl22}. Furthermore, by controlling experimental conditions, the single-layer sample of the SL-TL La$_3$Ni$_3$O$_7$ (La$_2$NiO$_4$/La$_4$Ni$_3$O$_{10}$)~\cite{Chen:jacs,Puphal:arxiv12,Wang:ic,Abadi:arxiv24} or SL-BL La$_5$Ni$_3$O$_{11}$ (La$_2$NiO$_4$/La$_3$Ni$_2$O$_7$)~\cite{,Li:prm24} nickelates have been also achieved already, showing that this procedure is realistic. Thus, our results provide clear predictions for future experiments on this specific material and other these promising Ni-oxide superconductors.

\noindent {\bf \\Methods\\}

\noindent {\small \bf DFT method\\}

In the present work, first-principles DFT calculations were performed using the Vienna {\it ab initio} simulation package (VASP) code with the projector augmented wave (PAW) method~\cite{Kresse:Prb,Kresse:Prb96,Blochl:Prb}, where the electronic correlations were considered by the generalized gradient approximation (GGA) and the Perdew-Burke-Ernzerhof (PBE) exchange potential~\cite{Perdew:Prl}. For the discussion of alternating stacking bilayer (BL) trilayer (TL) La$_7$Ni$_5$O$_{17}$, the atomic positions and crystal lattice were fully relaxed until the Hellman-Feynman force on each atom was smaller than $0.001$ eV/{\AA} for different pressures. Here, the plane-wave cutoff energy was set as $550$~eV and the $k$-point mesh was $20\times20\times3$ for the conventional structure. For the phonon spectrum of the P4mm (No. 99 ) and P4/mmm (No. 123) phases of the alternating BL-TL stacking La$_4$Ni$_3$O$_{17}$, a  $\sqrt2\times\sqrt2\times1$ supercell structure was used in the phonon calculation, by using the density functional perturbation theory approach~\cite{Baroni:Prl,Gonze:Pra1}, analyzed by the PHONONPY software in the primitive unit cell~\cite{Chaput:prb,Togo:sm}.  The evolution of possible phase transitions under pressure and structure at ambient conditions are not the main scope of our present work, thus we leave the issue of adding more pressures to future work. In addition to the standard DFT calculation, the maximally localized Wannier functions (MLWFs) method was also employed to fit the Ni $3d$ bands, by using the WANNIER90 packages~\cite{Mostofi:cpc}, to obtain the hopping matrices and crystal-field splittings. All the crystal structures were visualized with the VESTA code~\cite{Momma:vesta}.

\noindent {\small \bf Tight-binding method\\}
Based on the hopping matrices and crystal-field splittings obtained from MLWFs, we constructed a ten-band low-energy $e_g$-orbital BL plus TL tight-binding model, where the overall filling was considered as $n = 7.0$, corresponding to $1.4$ electrons per Ni site (for details of the origin of these numbers see main portion). The detailed hopping matrices can be found in an attached separate file. The Fermi energy is obtained by integrating the density of states for all $\omega$ until the number of electrons $n = 7$ is reached. Based on the obtained Fermi energy, a $4001\times4001$ $k$-mesh was used to calculate the Fermi surface.

\noindent {\small \bf RPA method\\}
To study superconducting pairing properties, the RPA method was used here, based on a perturbative weak-coupling expansion in the Hubbard interaction. It has been shown in many studies that this procedure captures the essence of the physics for Fe-based and Cu-based superconductors (see Ref.~\cite{Graser2009} as an example).
The full Hubbard model Hamiltonian for the BL plus TL discussed here, includes the kinetic energy and interaction terms, and it is written as $H = H_{\rm k} + H_{\rm int}$.
The electronic interaction portion of the Hamiltonian includes the standard same-orbital Hubbard repulsion $U$, the electronic repulsion $U'$ between electrons
at different orbitals, the Hund's coupling $J$, and the on-site inter-orbital electron-pair hopping terms ($J'$). Formally, it is given by:
\begin{eqnarray}
H_{\rm int}= U\sum_{i\gamma}n_{i \uparrow\gamma} n_{i \downarrow\gamma} +(U'-\frac{J}{2})\sum_{\substack{i\\\gamma < \gamma'}} n_{i \gamma} n_{i\gamma'} \nonumber \\
-2J \sum_{\substack{i\\\gamma < \gamma'}} {{\bf S}_{i,\gamma}}\cdot{{\bf S}_{i,\gamma'}}+J \sum_{\substack{i\\\gamma < \gamma'}} (P^{\dagger}_{i\gamma} P^{\phantom{\dagger}}_{i\gamma'}+H.c.),
\end{eqnarray}
where the standard relations $U'=U-2J$ and $J' = J$ are assumed, and $P_{i\gamma}$=$c_{i \downarrow \gamma} c_{i \uparrow \gamma}$.

In the multi-orbital RPA approach \cite{Kubo2007,Graser2009,Altmeyer2016,Romer2020}, the enhanced spin susceptibility is obtained from the bare susceptibility (Lindhart function) via $\chi_0({\bf q})$ as $\chi({\bf q}) = \chi_0({\bf q})[1-{\cal U}\chi_0({\bf q})]^{-1}$. Here, $\chi_0({\bf q})$ is an orbital-dependent susceptibility tensor and ${\cal U}$ is a tensor that contains the intra-orbital $U$ and inter-orbital $U'$ density-density interactions, the Hund's rule coupling $J$, and the pair-hopping $J'$ term. The pairing strength $\lambda_\alpha$ for channel $\alpha$ and the corresponding gap structure $g_\alpha({\bf k})$ are obtained from solving an eigenvalue problem of the form
\begin{eqnarray}\label{eq:pp}
	\int_{FS} d{\bf k'} \, \Gamma({\bf k -  k'}) g_\alpha({\bf k'}) = \lambda_\alpha g_\alpha({\bf k})\,,
\end{eqnarray}
where the momenta ${\bf k}$ and ${\bf k'}$ are on the FS and $\Gamma({\bf k - k'})$ contains the irreducible particle-particle vertex. In the RPA approximation, the dominant term entering $\Gamma({\bf k-k'})$ is the RPA spin susceptibility $\chi({\bf k-k'})$.

\noindent {\bf {\small \\ Data availability\\}} The input paramters for our TB calculations are available in the sperate file in the Supplementary information. Any additional data that support the findings of this study are available from the corresponding author upon request.
\noindent {\bf {\small \\ Code availability\\}} The Ab initio calculations are done with the code VASP.  Simulation RPA codes are available from the corresponding author upon reasonable request.

\noindent {\bf {\\Acknowledgements\\}}
\small {The work was supported by the U.S. Department of Energy (DOE), Office of Science, Basic Energy Sciences (BES), Materials Sciences and Engineering Division.}

\noindent {\bf {\\ Author contributions\\}} \small {Y.Z., L.-F.L., T.A.M., and E.D. designed the project. Y.Z., L.F.L., and T.A.M. carried out numerical calculations for DFT, the TB model, and RPA calculations. Y.Z., A.M., T.A.M., and E.D. wrote the manuscript. All co-authors provided useful comments and discussion on the paper.}

\noindent {\bf {\\Competing interests\\}} \small {The authors declare no competing interest.}

\noindent {\bf {\\ Additional information\\}}
Correspondence should be addressed to Ling-Fang Lin, Thomas A. Maier or Elbio Dagotto.

\noindent \small {{\bf Supplementary information} The online version contains supplementary material available at Commun. Phys. \href{https://doi.org/10.1038/xxxxx}{https://doi.org/10.1038/xxxxx.}}


\begin{references}
\bibitem{Sun:arxiv} Sun, H. {\it et al.} {Signatures of superconductivity near 80 K in a nickelate under high pressure} \href{https://doi.org/10.1038/s41586-023-06408-7}{\it Nature} {\textbf{621}, 493 (2023).}
\bibitem{Zhu:arxiv11} Zhu, Y.{\it et al.}  {J. Superconductivity in pressurized trilayer La$_4$Ni$_3$O$_{10-\delta}$ single crystals} \href{https://doi.org/10.1038/s41586-024-07553-3}{\it Nature} {\textbf{631} 531 (2024).}
\bibitem{Li:cpl} Li, Q., Zhang, Y.-J., Xiang, Z.-N., Zhang, Y., Zhu, X., \&  Wen, H.-H. {Signature of Superconductivity in Pressurized La$_4$Ni$_3$O$_{10}$} \href{https://doi.org/10.1088/0256-307X/41/1/017401}{\it Chinese Phys. Lett.} {\textbf{41}, 017401 (2024).}
\bibitem{LiuZhe:arxiv} Liu, Z. {\it et al.} {Electronic correlations and partial gap in the bilayer nickelate La$_3$Ni$_2$O$_7$} \href{https://doi.org/10.48550/arXiv.2307.02950}{\it arXiv} {2307.02950 (2023).}
\bibitem{Zhang:arxiv-exp} Zhang, Y. {\it et al.} {High-temperature superconductivity with zero resistance and strange-metal behaviour in La$_3$Ni$_2$O$_{7-\delta}$} \href{https://doi.org/10.1038/s41567-024-02515-y}{\it Nat. Phys.} {(2024).}
\bibitem{Hou:arxiv} Hou, J. {\it et al.} {Emergence of High-Temperature Superconducting Phase in Pressurized La$_3$Ni$_2$O$_7$ Crystals} \href{https://doi.org/10.1088/0256-307X/40/11/117302}{\it Chinese Phys. Lett.} {\textbf{40} 117302 (2023).}
\bibitem{Yang:arxiv09} Yang, J. {\it et al.} {Orbital-dependent electron correlation in double-layer nickelate La$_4$Ni$_3$O$_7$} \href{https://doi.org/10.1038/s41467-024-48701-7}{\it Nat. Commun.} {\textbf{15} 4373 (2024).}
\bibitem{Wang:arxiv9} Wang, G. {\it et al.} {Pressure-Induced Superconductivity In Polycrystalline La$_3$Ni$_2$O$_{7-\delta}$} \href{https://doi.org/10.1103/PhysRevX.14.011040}{\it Phys. Rev. X} {\textbf{14} 011040 (2024).}
\bibitem{Dong:arxiv12} Dong, Z. {\it et al.} {Visualization of Oxygen Vacancies and Self-doped Ligand Holes in La$_3$Ni$_2$O$_{7-\delta}$} \href{https://doi.org/10.1038/s41586-024-07482-1}{\it Nature} {\textbf{630} 847 (2024).}
\bibitem{Sakakibara:arxiv09} Sakakibara, H. {\it et al.} {Theoretical analysis on the possibility of superconductivity in the trilayer Ruddlesden-Popper nickelate La$_4$Ni$_3$O$_{10}$ under pressure and its experimental examination: Comparison with La$_3$Ni$_2$O$_7$} \href{https://doi.org/10.1103/PhysRevB.109.144511}{\it Phys. Rev. B} {\textbf{109} 125119 (2024).}
\bibitem{Zhang:arxiv11} Zhang, M. {\it et al.} {Superconductivity in trilayer nickelate La$_4$Ni$_3$O$_{10}$ under pressure} \href{https://doi.org/10.48550/arXiv.2311.07423} {\it arXiv}{2311.07423 (2023).}
\bibitem{Zhang:jmst} Zhang, M. {\it et al.} {Effects of pressure and doping on Ruddlesden-Popper phases La$_{\rm n+1}$Ni$_n$O$_{\rm 3n+1}$} \href{https://doi.org/10.1016/j.jmst.2023.11.011}{\it J. Mater. Sci. Technol.} {\textbf{185} 147 (2024).}
\bibitem{Whang:jac} Wang, L. {\it et al.} {Structure Responsible for the Superconducting State in La$_3$Ni$_2$O$_7$ at High-Pressure and Low-Temperature Conditions} \href{https://doi.org/10.1021/jacs.3c13094}{\it J. Am. Chem. Soc.} {\textbf{146} 7506 (2024).}
\bibitem{Li:arxiv24} Li, J. {\it et al.} {Pressure-driven right-triangle shape superconductivity in bilayer nickelate La$_3$Ni$_2$O$_7$} \href{https://doi.org/10.48550/arXiv.2404.11369}{\it arXiv}{2404.11369(2024).}
\bibitem{Zhang:nc20} Zhang, J. {\it et al.} {Intertwined density waves in a metallic nickelate} \href{https://doi.org/10.1038/s41467-020-19836-0}{\it Nat. Commun.} {\textbf{11}, 6003 (2020).}
\bibitem{Li:arxiv11} Li, J. {\it et al.} {Structural transition, electric transport, and electronic structures in the compressed trilayer nickelate La$_4$Ni$_3$O$_{10}$} \href{https://doi.org/10.1007/s11433-023-2329-x}{\it Sci. China Phys. Mech. Astron.} {\textbf{67}, 114703 (2024).}
\bibitem{Luo:prl23} Luo, Z., Hu, X., Wang, M., Wu, W., \& Yao, D.-X. {Bilayer Two-Orbital Model of La$_3$Ni$_2$O$_7$ under Pressure} \href{https://doi.org/10.1103/PhysRevLett.131.126001}{\it Phys. Rev. Lett.} {\textbf{131}, 126001 (2023).}
\bibitem{Zhang:prb23} Zhang, Y., Lin, L.-F., Moreo, A., \& Dagotto, E. {Electronic structure, dimer physics, orbital-selective behavior, and magnetic tendencies in the bilayer nickelate superconductor La$_3$Ni$_2$O$_7$ under pressure} \href{https://doi.org/10.1103/PhysRevB.108.L180510}{\it Phys. Rev. B} {\textbf{108}, L180510 (2023).}
\bibitem{Zhang:arxiv24} Zhang, Y., Lin, L.-F., Moreo, A., Maier, T. A. \& Dagotto, E. {Prediction of $s^{\pm}$-wave superconductivity enhanced by electronic doping in trilayer nickelates La$_4$Ni$_3$O$_{10}$ under pressure} \href{https://doi.org/10.48550/arXiv.2402.05285}{\it arXiv}{2402.05285 (2024).}
\bibitem{Li:nc} Li, H.; Zhou, {\it et al.} {Fermiology and electron dynamics of trilayer nickelate La$_4$Ni$_3$O$_{10}$} \href{https://doi.org/10.1038/s41467-017-00777-0}{\it Nat. Commun.} {\textbf{8}, 704 (2017).}
\bibitem{LaBollita:prb24} LaBollita, H., Kapeghian, J., Norman, M. R., \& Botana, A. S. {Electronic structure and magnetic tendencies of trilayer La$_4$Ni$_3$O$_{10}$ under pressure: Structural transition, molecular orbitals, and layer differentiation} \href{https://doi.org/10.1103/PhysRevB.109.195151}{\it Phys. Rev. B} {\textbf{109}, 195151 (2024).}
\bibitem{Tian:prb24} Tian, Y.-H., Chen, Y., Wang, J.-M., He, R.-Q., \& Lu, Z.-Y. {Correlation effects and concomitant two-orbital $s_{\pm}$-wave superconductivity in La$_3$Ni$_2$O$_7$ under high pressure} \href{https://doi.org/10.1103/PhysRevB.109.165154}{\rm Phys. Rev. B} {\textbf{109}, 165154 (2024).}
\bibitem{Wang:prb24} Wang, J.-X., Ouyang, Z., He, R-Q., \& Lu., Z.-Y. {Non-Fermi liquid and Hund correlation in La$_4$Ni$_3$O$_{10}$ under high pressure} \href{https://doi.org/10.1103/PhysRevB.109.165140}{\rm Phys. Rev. B} {\textbf{100}, 165140 (2024).}
\bibitem{Momma:vesta} Momma K., \& Izumi, F. {VESTA 3 for three-dimensional visualization of crystal, volumetric and morphology data} \href{https://doi.org/10.1107/S0021889811038970}{\rm J. Appl. Crystallogr.} {\textbf{44}, 1272 (2011).}
\bibitem{Zhang:prb24} Zhang, Y., Lin, L.-F., Moreo, A., Maier, T. A., \& Dagotto, {E. Electronic structure, magnetic correlations, and superconducting pairing in the reduced Ruddlesden-Popper bilayer La$_3$Ni$_2$O$_6$ nder pressure: Different role of $d_{3z^2-r^2}$ orbital compared with La$_3$Ni$_2$O$_7$} \href{https://doi.org/10.1103/PhysRevB.109.045151}{\rm Phys. Rev. B} {\textbf{109}, 045151 (2024).}
\bibitem{Chen:arxiv2024} Chen, X. {\it et al.} {Electronic and magnetic excitations in La$_3$Ni$_2$O$_7$} \href{https://doi.org/10.48550/arXiv.2401.12657}{\it arXiv}{2401.12657 (2024).}
\bibitem{Chen:prL24} Chen, K. {\it et al.} {Evidence of Spin Density Waves in high pressure La$_3$Ni$_2$O$_{7-\delta}$} \href{https://doi.org/10.1103/PhysRevLett.132.256503}{\it Phys. Rev. Lett.} {\textbf{132}, 256503 (2024).}
\bibitem{Xie:SB} Xie, T. {\it et al.} {Strong interlayer magnetic exchange coupling in La$_3$Ni$_2$O$_{7-\delta}$ revealed by inelastic neutron scattering} \href{https://doi.org/10.1016/j.scib.2024.07.030}{\it Science Bulletin} {(2024).}
\bibitem{Yang:prb23} Yang, Q.-G., Wang, D., \& Wang, Q.-H. {Possible $s_{\pm}$-wave superconductivity in La$_3$Ni$_2$O$_7$} \href{https://doi.org/10.1103/PhysRevB.108.L140505}{\it Phys. Rev. B} {\textbf{108}, L140505 (2023).}
\bibitem{Zhang:nc24} Zhang, Y., Lin, L.-F., Moreo, A., Maier, T. A., \& Dagotto, E. {Structural phase transition, $s^{\pm}$-wave pairing, and magnetic stripe order in bilayered superconductor La$_3$Ni$_2$O$_7$ under pressure} \href{https://doi.org/10.1038/s41467-024-46622-z}{\it Nat. Commun.} {\textbf{15}, 2470 (2024).}
\bibitem{Liu:prl23} Liu, Y.-B., Mei, J.-W., Ye, F., Chen, W.-Q., \& Yang, F. {$s^{\pm}$-Wave Pairing and the Destructive Role of Apical-Oxygen Deficiencies in La$_3$Ni$_2$O$_7$ under Pressure} \href{https://doi.org/10.1103/PhysRevLett.131.236002}{\it Phys. Rev. Lett.} {\textbf{131}, 236002 (2023).}
\bibitem{Liao:prb23} Liao, Z. {\it et al.} {Electron correlations and superconductivity in La$_3$Ni$_2$O$_7$ under pressure tuning} \href{https://doi.org/10.1103/PhysRevB.108.214522}{\it Phys. Rev. B} {\textbf{108}, 214522 (2023).}
\bibitem{Qu:prl} Qu, X.-Z. {\it et al.} {Bilayer ${t-J-J_{\perp}}$ Model and Magnetically Mediated Pairing in the Pressurized Nickelate La$_3$Ni$_2$O$_7$} \href{https://doi.org/10.1103/PhysRevLett.132.036502}{\it Phys. Rev. Lett.} {\textbf{132}, 036502 (2024).}
\bibitem{Yang:arxiv24} Yang, Q.-G., Jiang, K.-Y., Wang, D., Lu, H.-Y., \& Wang, Q.-H. {Effective model and $s_{\pm}$-wave superconductivity in trilayer nickelate La$_4$Ni$_3$O$_{10}$} \href{https://doi.org/10.1103/PhysRevB.109.L220506}{\it Phys. Rev. B} {\textbf{109}, L220506 (2024).}
\bibitem{Zhang:prb23-2} Zhang, Y., Lin, L.-F., Moreo, A., Maier, T. A., \& Dagotto, E. {Trends in electronic structures and $s_{\pm}$-wave pairing for the rare-earth series in bilayer nickelate superconductor $R_3$Ni$_2$O$_7$} \href{https://doi.org/10.1103/PhysRevB.108.165141}{\it Phys. Rev. B} {\textbf{108}, 165141 (2023).}
\bibitem{Geisler:qm} Geisler, B., Hamlin, J. J., Stewart, G. R., Hennig R. G., \& Hirschfeld, P. J. {Structural transitions, octahedral rotations, and electronic properties of $A_3$Ni$_2$O$_7$  rare-earth nickelates under high pressure} \href{https://doi.org/10.1038/s41535-024-00648-0}{\it npj Quantum Mater.} {\textbf{9}, 38 (2024).}
\bibitem{Lu:prl} Lu, C., Pan, Z., Yang, F., \& Wu, C. {Interlayer-Coupling-Driven High-Temperature Superconductivity in La$_3$Ni$_2$O$_7$ under Pressure} \href{https://doi.org/10.1103/PhysRevLett.132.146002}{\it Phys. Rev. Lett.} {\textbf{132}, 146002 (2024).}
\bibitem{Cao:prb23} Cao, Y., \&  Yang, Y.-f. {Flat bands promoted by Hund's rule coupling in the candidate double-layer high-temperature superconductor La$_3$Ni$_2$O$_7$ under high pressure} \href{https://doi.org/10.1103/PhysRevB.109.L081105}{\it Phys. Rev. B} {\textbf{109}, L081105 (2024).}
\bibitem{Kim:prr22} Kim, H. {\it et al.} {Sr$_2$NiO$_4$/Sr$_3$Ir$_2$O$_7$ superlattice for a model two-dimensional quantum Heisenberg antiferromagnet} \href{https://doi.org/10.1103/PhysRevResearch.4.013229}{\it Phys. Rev. Research} {\textbf{4}, 013229 (2022).}
\bibitem{Gong:prl22} Gong, D. {\it et al.} {Reconciling Monolayer and Bilayer $J_{\rm eff} = 1/2$ quare Lattices in Hybrid Oxide Superlattice} \href{https://doi.org/10.1103/PhysRevLett.129.187201}{\it Phys. Rev. Lett.} {\textbf{129}, 187201 (2022).}
\bibitem{Zhai:nc14} Zhai, X. {\it et al.} {Correlating interfacial octahedral rotations with magnetism in (LaMnO$_{3+\delta}$)$_N$/(SrTiO$_3$)$_N$ superlattices} \href{https://doi.org/10.1038/s41467-024-46622-z}{\it Nat. Commun.} {\textbf{15}, 2470 (2024).}
\bibitem{Chen:nl} Chen, B. {\it et al.} {Spatially Controlled Octahedral Rotations and Metal-Insulator Transitions in Nickelate Superlattices} \href{https://doi.org/10.1021/acs.nanolett.0c03850}{\it Nano Lett.} {\textbf{21}, 1295 (2021).}
\bibitem{Yang:nl} Yang, C. {\it et al.} {Thickness-Dependent Interface Polarity in Infinite-Layer Nickelate Superlattices} \href{https://doi.org/10.1021/acs.nanolett.3c00192}{\it Nano Lett.} {\textbf{23}, 3291  (2023).}
\bibitem{Chen:jacs} Chen, X. {\it et al.} {Polymorphism in the Ruddlesden-Popper Nickelate La$_3$Ni$_2$O$_7$: Discovery of a Hidden Phase with Distinctive Layer Stacking} \href{https://doi.org/10.1021/jacs.3c14052}{\it J. Am. Chem. Soc.} {\textbf{146}, 23640 (2024).}
\bibitem{Wang:ic} Wang, H.; Chen, L.; Rutherford, A.; Zhou, H.; Xie, W. Long-Range Structural Order in a Hidden Phase of Ruddlesden-Popper Bilayer Nickelate La$_3$Ni$_2$O$_7$ \href{https://doi.org/10.1021/acs.inorgchem.3c04474} {Inorg. Chem. \textbf{63}, 5020 (2024).}
\bibitem{Puphal:arxiv12} Puphal, P. {\it et al.} {Unconventional crystal structure of the high-pressure superconductor La$_3$Ni$_2$O$_7$} \href{https://doi.org/10.48550/arXiv.2312.07341}{\it arXiv} {2312.07341 (2023).}
\bibitem{Abadi:arxiv24} Abadi, S. N. {\it et al.} {Electronic structure of the alternating monolayer-trilayer phase of La$_3$Ni$_2$O$_7$} \href{https://doi.org/10.48550/arXiv.2402.07143}{\it arXiv} {2402.07143 (2024).}
\bibitem{Li:prm24} Li, F. {\it et al.} {Design and synthesis of three-dimensional hybrid Ruddlesden-Popper nickelate single crystals} \href{https://doi.org/10.1103/PhysRevMaterials.8.053401}{\it Phys. Rev. Materials} {\textbf{8}, 053401 (2024).}
\bibitem{Kresse:Prb} Kresse, G., \& Hafner J. {Ab initio molecular dynamics for liquid metals.} \href{https://doi.org/10.1103/PhysRevB.47.558}{\it Phys. Rev. B} {\textbf{47}, 558 (1993).}
\bibitem{Kresse:Prb96} Kresse, G., \& Furthm\"{u}ller, J. {Efficient iterative schemes for ab initio total-energy calculations using a plane-wave basis set.} \href{https://doi.org/10.1103/PhysRevB.54.11169}{\it Phys. Rev. B} {\textbf{54}, 11169 (1996).}
\bibitem{Blochl:Prb} Bl\"{o}chl, P. E. {Projector augmented-wave method.} \href{https://doi.org/10.1103/PhysRevB.50.17953}{Phys. Rev. B} {\textbf{50}, 17953 (1994).}
\bibitem{Perdew:Prl} Perdew, J. P., K. Burke, \& Ernzerhof, M. {Generalized Gradient Approximation Made Simple.} \href{https://doi.org/10.1103/PhysRevLett.77.3865}{\it Phys. Rev. Lett.} {\textbf{77}, 3865 (1996).}
\bibitem{Sakakibara:prl24} Sakakibara, H., Kitamine, N., Ochi, M., \& Kuroki, K. {Possible High $T_c$ Superconductivity in La$_3$Ni$_2$O$_7$ under High Pressure through Manifestation of a Nearly Half-Filled Bilayer Hubbard Model} \href{https://doi.org/10.1103/PhysRevLett.132.106002}{\it Phys. Rev. Lett.} {\textbf{132}, 106002 (2024).}
\bibitem{Zhang:1313} Zhang, Y., Lin, L.-F., Moreo, A., Maier, T. A.,  \& Dagotto, E. {Electronic structure, self-doping, and superconducting instability in the alternating single-layer trilayer stacking nickelates La$_3$Ni$_2$O$_7$} \href{https://doi.org/10.48550/arXiv.2404.16600}{\it arXiv} {2404.16600 (2024).}
\bibitem{Baroni:Prl} BBaroni, S., Giannozzi, P. \& Testa, A.  {Green's-function approach to linear response in solids} \href{https://doi.org/10.1103/PhysRevLett.58.1861}{\it Phys. Rev. Lett.} {\textbf{58}, 1861 (1987).}
\bibitem{Gonze:Pra1} Gonze, X. {Perturbation expansion of variational principles at arbitrary order} \href{https://doi.org/10.1103/PhysRevA.52.1086}{Phys. Rev. A} {\textbf{52}, 1086 (1995).}
\bibitem{Gonze:Pra2} Gonze, X. {Adiabatic density-functional perturbation theory} \href{https://doi.org/10.1103/PhysRevA.52.1096}{\it Phys. Rev. A} {\textbf{52}, 1096 (1995).}
\bibitem{Chaput:prb} Chaput, L., Togo, A., Tanaka, I. \& Hug, G. {Phonon-phonon interactions in transition metals} \href{https://doi.org/10.1103/PhysRevB.84.094302}{\it Phys. Rev. B} {\textbf{84}, 094302 (2011).}
\bibitem{Togo:sm} Togo, A. Tanaka, \& I. {First principles phonon calculations in materials science} \href{https://doi.org/10.1016/j.scriptamat.2015.07.021}{\it Scr. Mater.} {\textbf{108}, 1 (2015).}
\bibitem{Nomura:rpp} Nomura, Y., \& Arita, R. {Superconductivity in infinite-layer nickelates} \href{https://doi.org/10.1088/1361-6633/ac5a60}{\it Rep. Prog. Phys.} {\textbf{85}, 052501  (2022).}
\bibitem{Zhang:prb20} Zhang, Y. {\it et al.} {Similarities and differences between nickelate and cuprate films grown on a SrTiO$_3$ substrate} \href{https://doi.org/10.1103/PhysRevB.102.195117}{\it Phys. Rev. B} {\textbf{102}, 195117 (2020).}
\bibitem{Mostofi:cpc} Mostofi, A.~A. {\it et al.} {Wannier90: A tool for obtaining maximally-localised wannier functions.} \href {https://doi.org/10.1016/j.cpc.2007.11.016} {\it Comput. Phys. Commun.} { \textbf {178}, 685-699 (2008)}.
\bibitem{Romer2020} R{\o}mer, A. T., {\it et al.} {Pairing in the two-dimensional Hubbard model from weak to strong coupling} \href{https://doi.org/10.1103/PhysRevResearch.2.013108}{\it Phys. Rev. Res.} {\textbf{2}, 13108 (2020).}
\bibitem{Kubo2007} Kubo, K., {Pairing symmetry in a two-orbital Hubbard model on a square lattice} \href{https://doi.org/10.1103/PhysRevB.75.224509}{\it Phys. Rev. B} {\textbf{8}, 224509 (2007).}
\bibitem{Graser2009}  Graser, S., Maier, T. A., Hirschfeld, P. J., \&  Scalapino, D. J., {Near-degeneracy of several pairing channels in multiorbital models for the Fe pnictides} \href{https://doi.org/10.1088/1367-2630/11/2/025016}{\it New J. Phys.} {\textbf{11}, 25016 (2009).}
\bibitem{Altmeyer2016} Altmeyer, M., {\it et al.} {Role of vertex corrections in the matrix formulation of the random phase approximation for the multiorbital Hubbard model} \href{https://doi.org/10.1103/PhysRevB.94.214515}{\it Phys. Rev. B} {\textbf{94}, 214515 (2016).}
\bibitem{Lin:prl21}  Lin, L.-F., Zhang, Y., Alvarez, G., Moreo, A. \& Dagotto, E. {Origin of insulating ferromagnetism in iron oxychalcogenide Ce$_2$O$_2$FeSe$_2$.} \href{https://doi.org/10.1103/PhysRevLett.127.077204} {\it Phys. Rev. Lett.} { \textbf {127}, 077204 (2021)}.
\bibitem{Lin:cp} Lin, L.-F. {\it et al.} {Stability of the interorbital-hopping mechanism for ferromagnetism in multi-orbital Hubbard models} \href{https://doi.org/10.1038/s42005-023-01314-w}{\it Commun. Phys.} {\textbf{6}, 199 (2023).}
\end{references}
\end{document}